\documentstyle[preprint,prl,aps]{revtex}
\input{epsf}

\begin{document}

\title{Parametrically excited surface waves in magnetic fluids:
observation of domain structures}

\date{\today}

\author{Thomas Mahr and Ingo Rehberg}
\address{ Institut f\"ur Experimentelle Physik,
          Otto--von--Guericke--Universit\"at,
          Postfach 4120,
          D-39016 Magdeburg,
          Germany}

%\author{Ingo Rehberg}
%\address{ Institut f\"ur Experimentelle Physik,
%          Otto--von--Guericke--Universit\"at,
%          Postfach 4120,
%          D-39016 Magdeburg,
%          Germany}

\maketitle

\begin{abstract}

Observations of parametrically excited surface waves in a magnetic fluid
are presented. Under the influence of a magnetic field these waves have a
non--monotonic dispersion relation, which leads to a richer behavior than
in ordinary liquids. We report observation of three novel effects, namely:
i)~domain structures, ii)~oscillating defects and iii)~relaxational phase
oscillations.

\end{abstract}


%\section{Introduction}
%======================

``Nam caelo terras et terris abscidit undas et liquidum spisso secrevit ab
aere caelum''\cite{Ovid}. Ovid describes here a vision of the origin of
the earth, arising from an unstructured ground state. The idea appears
amazingly modern: A physicist would use the term ``spontaneous
symmetry--breaking'' to describe such a process, and his vision of the
origin of the universe is apparently similar. Spontaneous formation of
patterns from unstructured ground states are investigated systematically
within the field of nonequilibrium pattern formation \cite{CrossHohenberg}.
The most popular example are surface waves generated by wind blowing over
water. Such waves have to be excited in a spatially homogenous manner ---
e.~g.~via parametric excitation generated by vertical vibration --- in
order to allow for spontaneous symmetry--breaking processes
\cite{FaradayLiteratur}. 

A separation of the surface wave into different phases has recently been
predicted for parametrically excited surface waves \cite{Riecke97}. This
intriguing result is due to a non--monotonic dispersion relation for surface
waves, which means that up to three different wave numbers can be excited
with one single driving frequency. The ensuing competition between these
different wave numbers is the reason for a phase separation in the form of
domain structures. A magnetic fluid is an experimental system with a
non--monotonic dispersion relation for surface waves \cite{MaRe95}. This
Letter reports the first experimental observation of the theoretically
predicted domain structures using that material. Furthermore, two novel
dynamic states caused by the nonlinear interaction of the surface waves
are described, which are presumably due to the unusual dispersion relation
of that complex fluid.

%\section{Experimental Setup}
%===========================

\begin{figure}[h]

\epsfxsize=15cm
\epsfbox{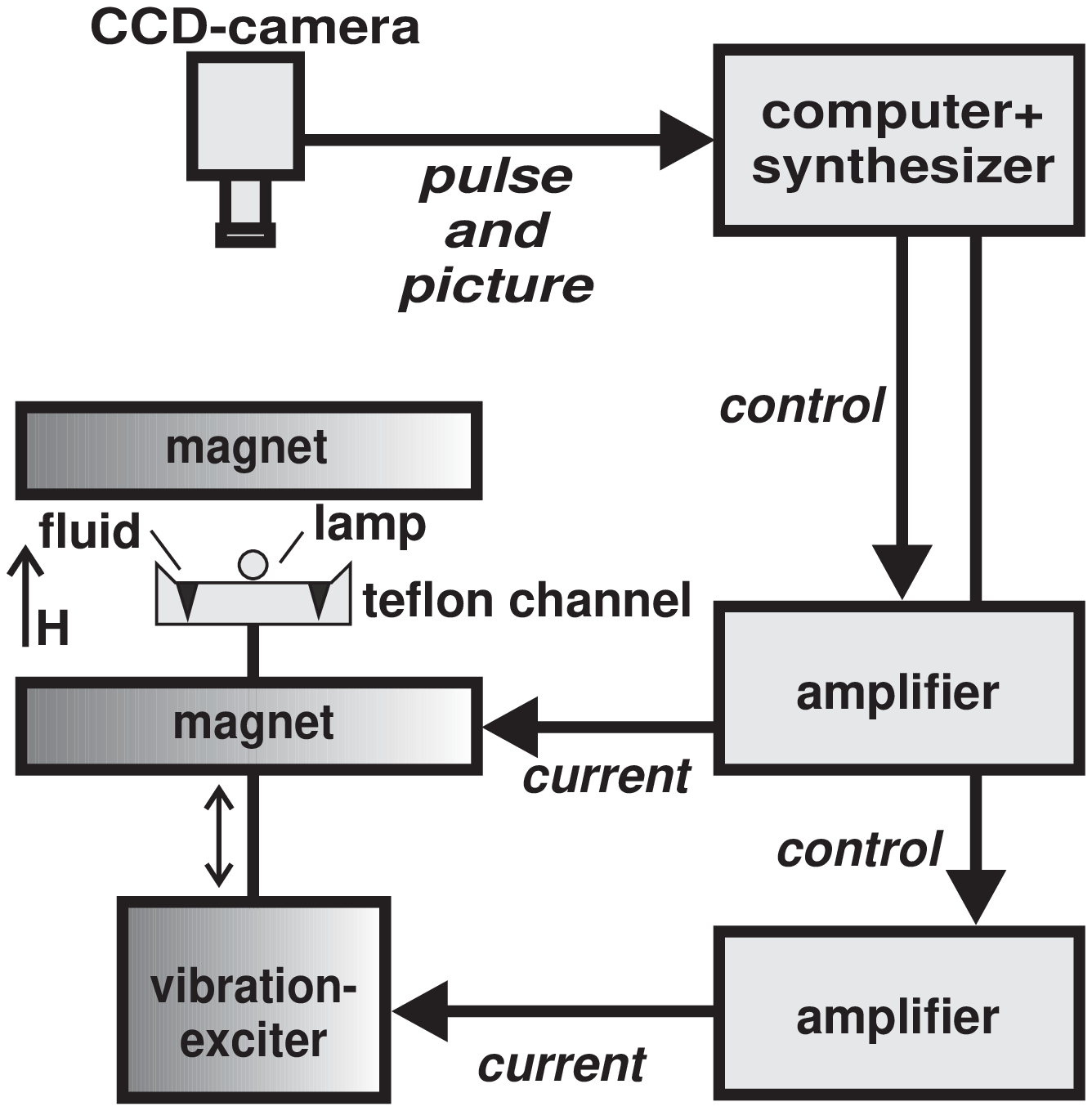}
\caption {\label{setup} Experimental setup. }
\end{figure}

The experimental setup is shown in Fig.~\ref{setup}. A V--shaped circular
channel was machined into a Teflon dish. The channel has a diameter of
60\,mm, a depth of 5\,mm and an upper width of 4\,mm, which is smaller
than the typical wavelength in the experiment \cite{MaRe95}. The upper
part of the channel has a slope of $15^{\circ}$ to the horizontal in order
to suppress surface waves in radial direction. The border of the dish has a
slope of $45^{\circ}$, and is used as a screen, where the shadow of the
surface waves of the fluid is projected. A light bulb placed in the center
of the channel casts a shadow of the surface on the screen.

The channel is filled with the magnetic fluid EMG 909, Lot.\,No.~F081996A
(Ferrofluidics). The properties of the fluid are: density $\rho =
1020$\,kgm$^{-3}$, surface tension $\sigma = 26.5 \cdot
10^{-3}$\,Nm$^{-1}$, initial magnetic permeability $\mu = 1.8$, magnetic
saturation $ M_S = 1.6 \cdot 10^4$\,Am$^{-1}$ and dynamic viscosity $
\eta= 6 \cdot 10^{-3}$\,Nsm$^{-2}$. The channel is placed in the center of
a pair of Helmholtz--coils (Oswald--Magnetfeldtechnik), with an inner
diameter of 40\,cm. Each coil consists of 474 windings of flat copper wire
with a width of 4.5\,mm and thickness 2.5\,mm. A current of about 10\,A is
supplied by a linear amplifier (fug NLN 5200 M-260). The magnetic field is
monitored by means of a hall probe (Group 3 DTM-141 Digital Teslameter)
located near the surface of the channel.

Control and analysis of the experiment is done with a 90\,MHz Pentium--PC,
equipped with a $512 \times 512$ 8--bit frame grabber (Data Translation
DT2853), and a synthesizer--board (WSB-10). The sine--signal of the
synthesizer--board is amplified by a linear amplifier (Br\"ul \& Kjaer
Power Amplifier Type 2712) and controls the oscillation of the
vibration--exciter (Br\"ul \& Kjaer PM Vibration Exciter Type 4808), which
drives the channel in vertical direction. For sufficiently large driving
amplitudes the flat surface becomes unstable and standing waves ensue. 

\begin{figure}[h]
\epsfxsize=15cm
\epsfbox{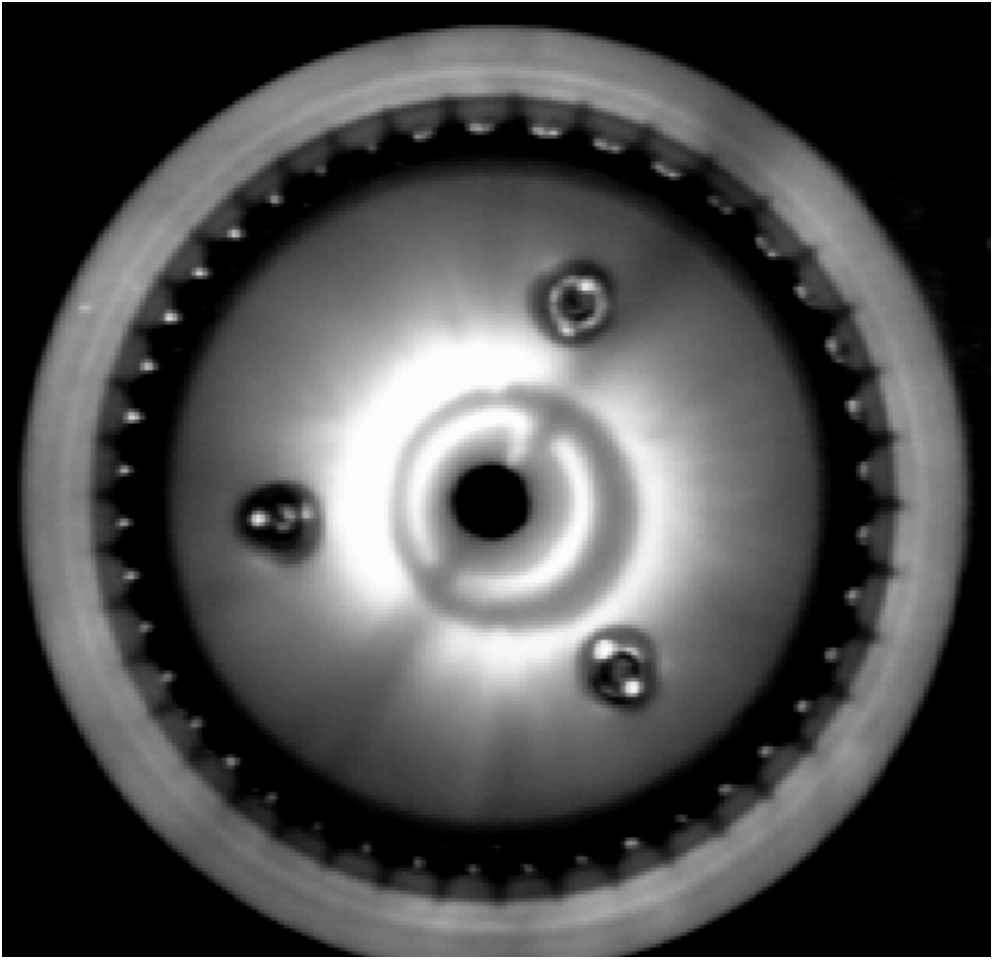}
\caption {\label{photo} Snapshot of the channel. The
shadows of the standing waves are visible. In this image there are two
domains of
different wave numbers, a longer wave in the upper right quadrant of 
the channel and a smaller wave in the rest of the channel. The magnetic 
field is $H = 0.96\,H_C$, wave number at the onset of the static 
instability $k_C = 56$.}
\end{figure}

The shadow of the standing waves is detected by a CCD--camera (Philips LDH
0600/00) placed 100 cm above the center of the channel. The camera
operates in the interlaced mode at 50\,Hz using an exposure time of 40\,ms
and supplies the frame grabber with images. A typical image is shown in
Fig.~\ref{photo}. One can see the bulb in the center of the dish. A screen
on the top of the bulb avoids direct light emission into the camera. The
real channel is only seen as a black ring in the image, but the shadow of
the surface waves can be seen clearly.

To observe the time evolution of the surface we use a phase--locked
technique between the driving and the sampling, where the
synthesizer--board is triggered by the 64\,$\mu s$ signal of the
horizontal synchronization of the camera \cite{MaRe95}. This
phase--locking enables us to analyze the dynamics of the fluid surface in
a Poincare section, where the sampling phase is chosen to observe maximum
amplitude of the standing waves. The data reduction of the
two--dimensional camera image to one line is described in
Ref.~\cite{MaRe95}.

%\section{Experimental Results} %==============================

In Ref.~\cite{Riecke97} it is suggested, that one could obtain domains of
coexisting wave numbers in a fluid with non--monotonic dispersion relation
for surface waves by quenching. In our system three parameters could be
quenched: the driving frequency, the amplitude, and the magnetic field.
The last one is most convenient experimentally. Therefore, in our
measurements we fix the driving frequency $f_D$ and start with a magnetic
field of $H=0.85$\,$H_C$, where $H_C$ is the critical field for the onset
of the Rosensweig--Instability \cite{RosenBuch}, where the flat surface is
unstable in a static magnetic field. Then, the mechanical driving is
increased to a value $10 \%$ above the onset of standing waves, which
oscillate at a frequency $f=f_D/2$. For fixed driving frequency and
amplitude the magnetic field $H$ serves as the control parameter. We
perform jumps of $H$ to quench the system into a bistable regime. The
jumps are random since there is no prediction for the parameter values
where the domains should arise. 

\begin{figure}[h]
\hspace{-1cm}
\epsfxsize=17cm
\epsfbox{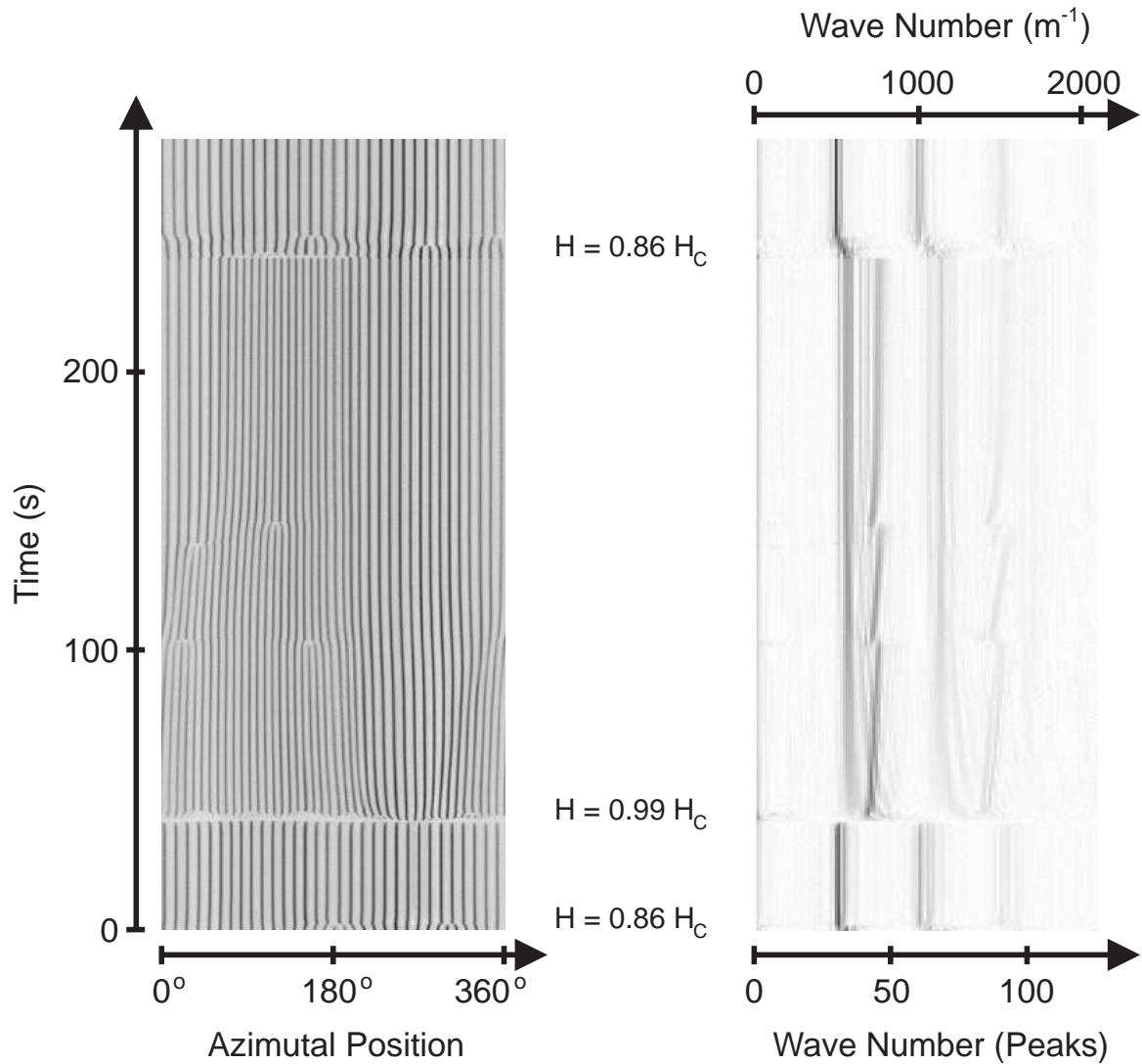}
\caption {\label{dom1} Observation of domains for $H<H_C$ at
$f_D=20.8$\,Hz. 
Left diagram: the surface is observed at constant phase. Black parts
correspond to wave crests, whereas white parts correspond to wave
troughs or a flat surface. Right diagram: the corresponding spectrum
is shown, obtained by a FFT. Between both diagrams the value of the
magnetic field is indicated. One can clearly see that domains of
different wavelengths exist at $H=0.99$\,$H_C$, $H_C=19000$\,$Am^{-1}$.}
\end{figure}

A camera snapshot of the channel with domains of excited surface waves
with different wavelengths is shown in Fig.~\ref{photo}. In the upper
right quadrant the wavelength is larger than it is in the rest of the
channel. From images like this we extract the amplitude of the standing
wave as a function of azimutal position and time \cite{MaRe95}. One
example, measured at $f_D=20.8$\,Hz, is presented in Fig.~\ref{dom1}. The
left diagram shows the time evolution of the surface along the azimutal
position at a constant phase. Black parts in this diagram correspond to
wave crests, where white parts correspond to wave troughs or a flat
surface. The right diagram presents the Fourier--spectrum obtained by a
one--dimensional Fast Fourier Transformation \cite{NumRec}. The scale at
the top gives the wave number $k_{SI}$ in SI--units, while the bottom
scale indicates the modes $k$ in the Fourier--space. For a homogeneous
system $k$ is the number of waves in the channel. At $t=0$ the magnetic
field is $H=0.86$\,$H_C$, and steady stable waves are present. The
standing wave is homogeneous, 31 wave crests are counted. A jump to
$H=0.99$\,$H_C$ leads to a flat surface almost instantaneously, and
subsequently to the creation of two domains with different wave numbers.
At $t=200$\,s the two wave numbers are $k_1=34$ and $k_2=46$. The two
domains can be seen in the space--time--diagram left. The
Fourier--spectrum right indicates the two corresponding modes. A jump back
to $H=0.86$\,$H_C$ leads to the destruction of the two domains. Only one
homogeneous wave number $k=30$ results. The dynamics is similar to that of
the first transition only in the short wave domain: the amplitude vanishes
almost instantaneously. The long wave domain on the other hand reorganizes
comparatively slowly. This is the first experimental demonstration of the
existence of domain structures, which stem from the non--monotonic
dispersion relation for surface waves in a ferrofluid. This phenomenon was
predicted in Ref.~\cite{Riecke97}.   

\begin{figure}[h]
\hspace{-1cm}
\epsfxsize=17cm
\epsfbox{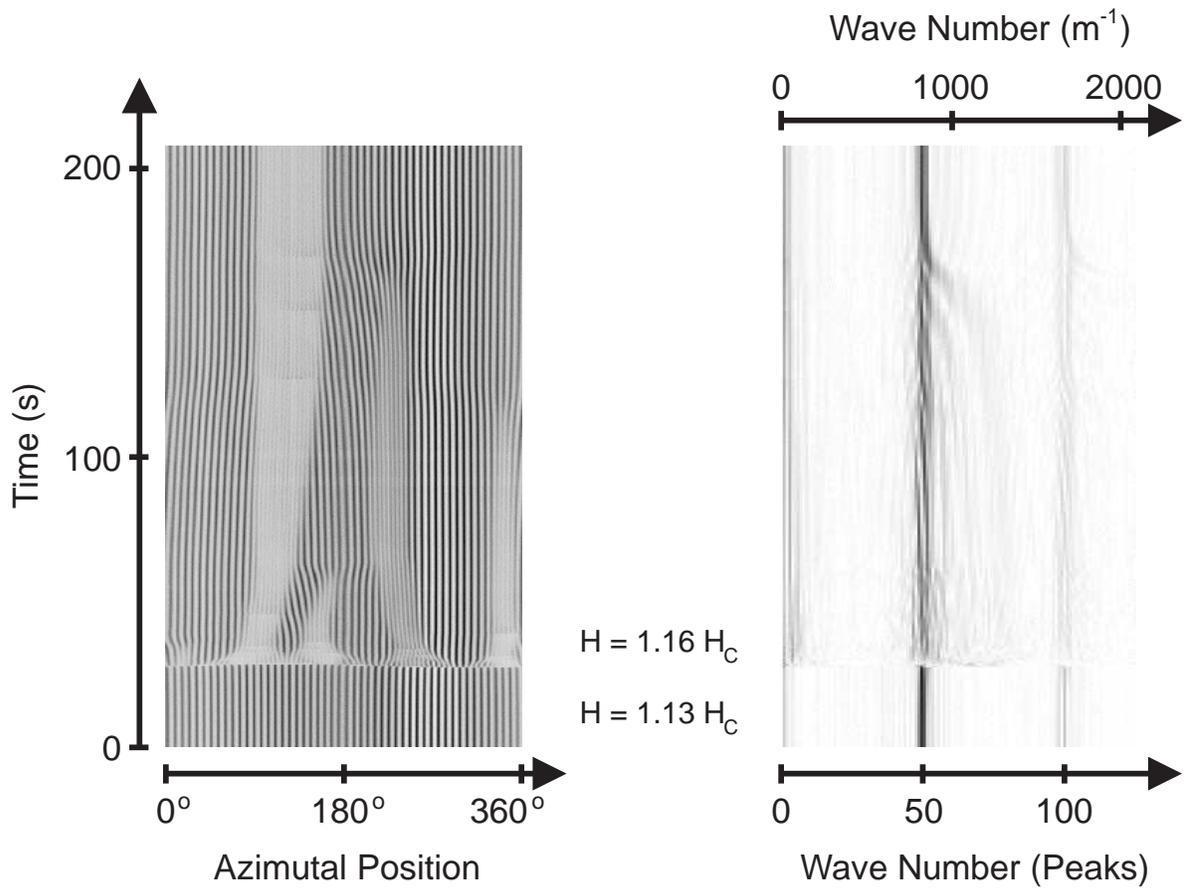}
\caption {\label{dom2} Observation of domains for $H>H_C$ at
$f_D=20.8$\,Hz. 
See figure capture of Fig.~\ref{dom1}.}
\end{figure}

Domains of coexisting wave numbers are not only observed for magnetic
fields below $H_C$, where this theory should be applicable, but also for
higher values of the magnetic field. A corresponding observation is
presented in Fig.~\ref{dom2}. Here, the driving frequency again is
$20.8$\,Hz and the magnetic field $H=1.13$\,$H_C$. A jump of the field to
an even higher value of $H=1.16$\,$H_C$ again induces domains, and the
dynamics of the transition is similar to that observed at smaller field.
The wave numbers in the steady state are now very different, $k_1=51$ and
$k_2=87$. The ratio of the two wave numbers is larger than for
Fig.~\ref{dom1}, and the amplitude modulation is even stronger, thus the
waves in the $k_2=87$ -- domain are hardly visible in the plot.

\begin{figure}[h]
\hspace{-1cm}
\epsfxsize=17cm
\epsfbox{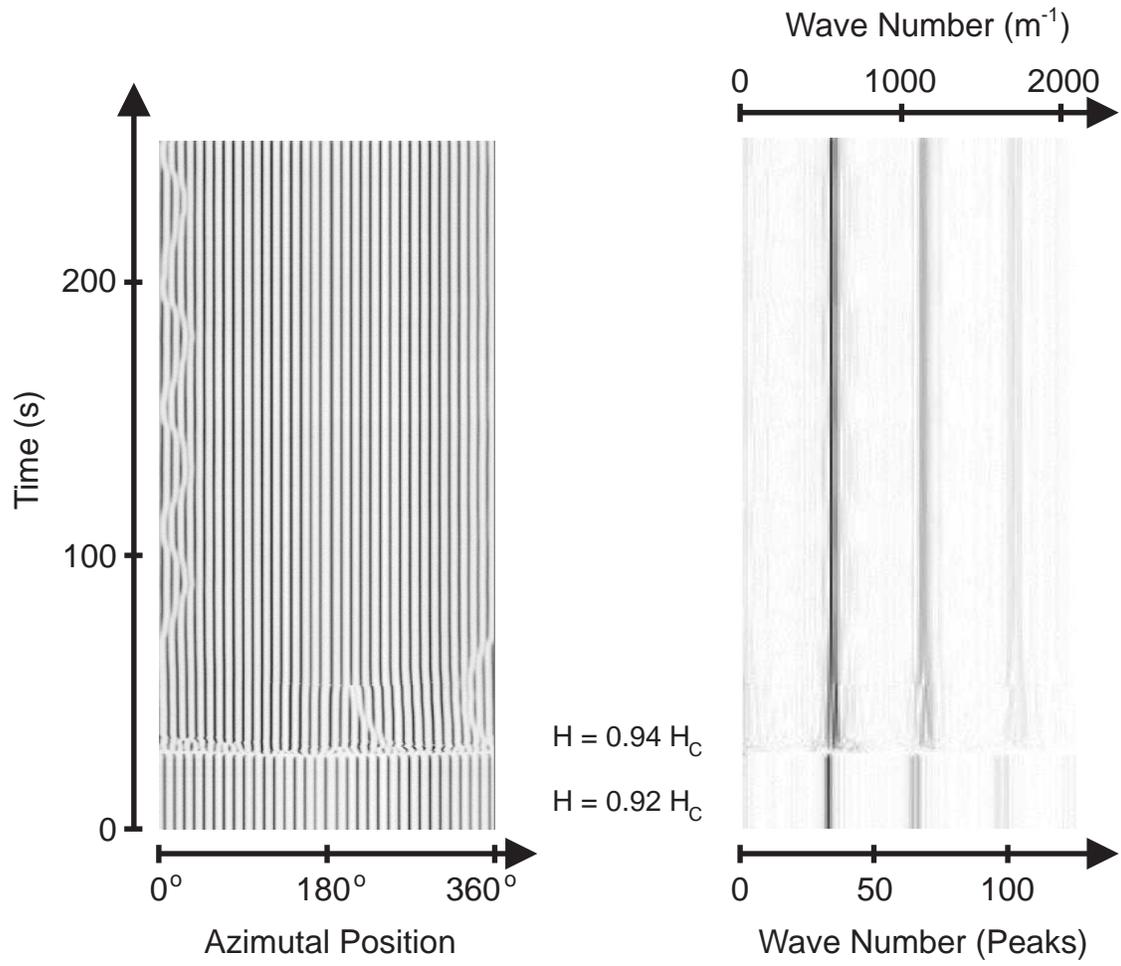}
\caption {\label{osz} Observation of oscillating defects at
$f_D=20.8$\,Hz. 
See figure capture of Fig.~\ref{dom1}.}
\end{figure}

The domains presented so far are in principal explained by theoretical
considerations. In addition we would like to present two novel oscillatory
modes, which go beyond that theory. The first example is shown in
Fig.~\ref{osz} starting at a homogeneous state at wave number $k=33$ and
$H=0.92$\,$H_C$ the field is increased to $H=0.94$\,$H_C$. The resulting
state is now time--depending, a defect occurs, oscillating in its position
with a period of about 45\,s. In addition other states have been observed,
where the oscillation amplitude is modulated in time.

\begin{figure}[h]
\hspace{-1cm}
\epsfxsize=17cm
\epsfbox{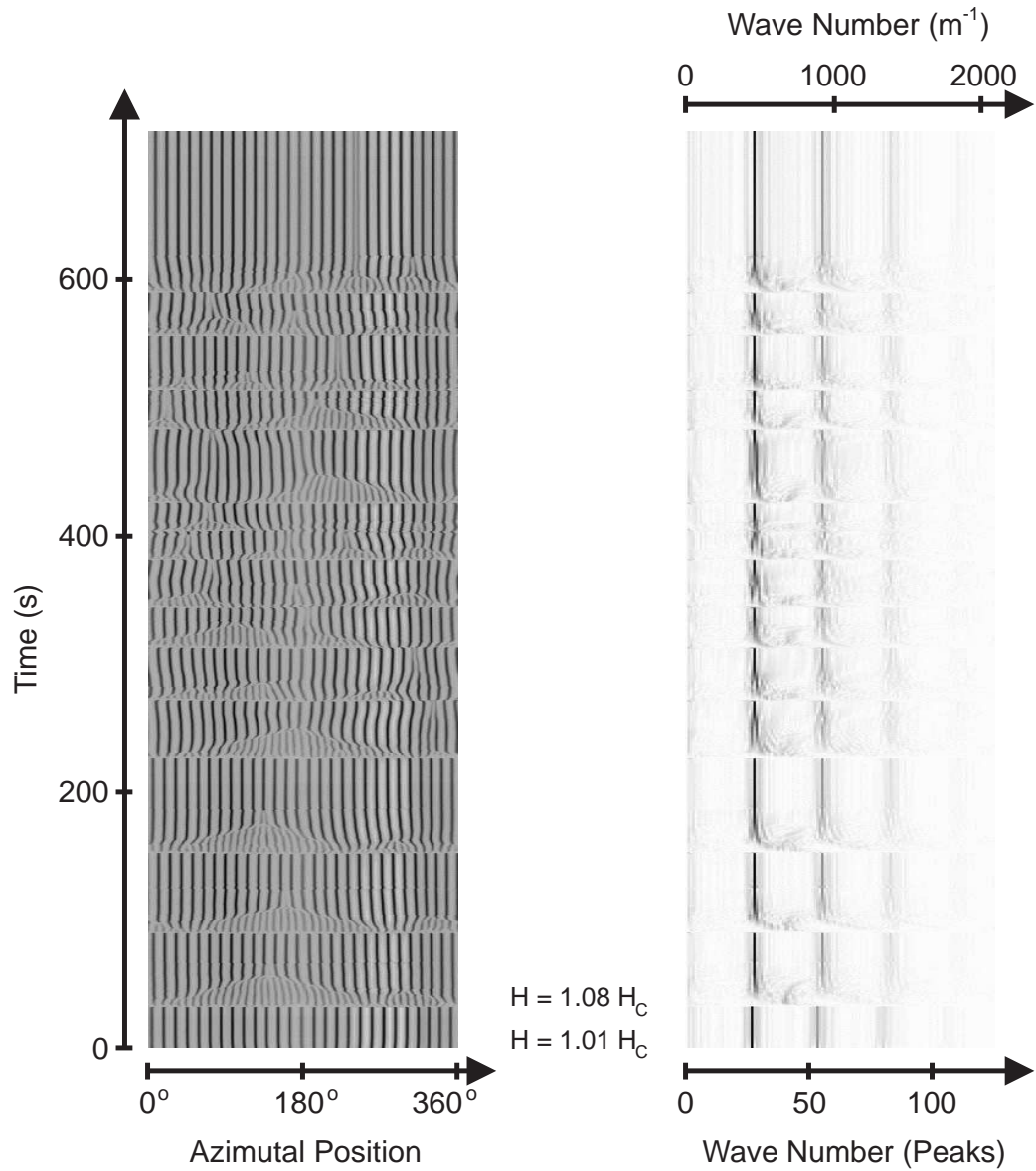}
\caption {\label{cascade} Observation of relaxational phase oscillations.
No external parameter is changed for $t>40s$. The driving frequency is 
$f_D=17.8$\,Hz. 
See figure capture of Fig.~\ref{dom1}.}
\end{figure}

Another time dependent state, named relaxational phase oscillation, is
presented in Fig.~\ref{cascade}. Here we start from homogeneous state with
wave number $k=27$ at the magnetic field $H=1.01$\,$H_C$. Increasing the
field to $H=1.08$\,$H_C$ leads to a time--dependent state with an
interesting dynamics. The first step consists of a fast decay of the
amplitude. A short wave subsequently grows, which is unstable leading to
the destruction of a few wave crests. The resulting homogeneous state with
a smaller wave number is again unstable. It leads to a sudden crash of the
standing wave amplitude and the cycle repeats. The oscillating behavior of
the phase with its fast and slow time scales is very reminiscent of a
relaxation oscillation.

\begin{figure}[h]
\epsfxsize=15cm
\epsfbox{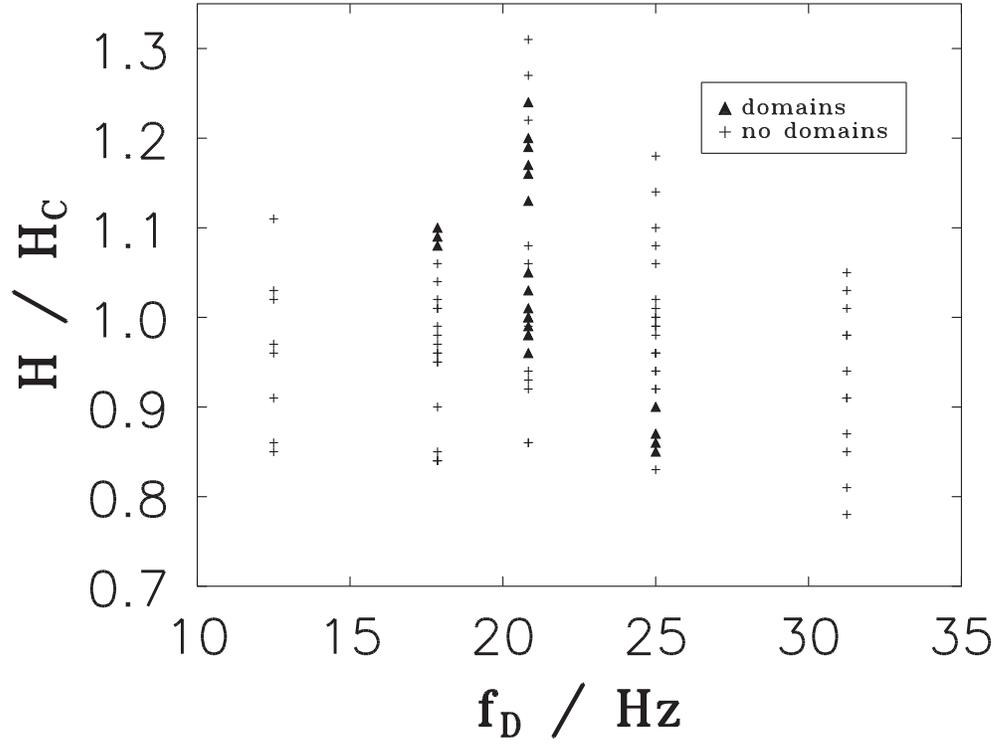}
\caption {\label{phase} Phase--diagram of coexisting domains with
different wave numbers as function of the driving frequency $f_D$
and the magnetic field $H$.}
\end{figure}

The oscillations presented here are two examples of many different
oscillatory modes. Systematic measurements have been made to investigate
the range of existence of domains and the phase diagram is presented in
Fig.~\ref{phase}. Apparently domains are created near $H=H_C$ for a
driving frequency of 20\,Hz. This corresponds to 10\,Hz of the
parametrically driven surface waves, which oscillate at half the driving
frequency. 10\,Hz is approximately the frequency, where the non--monotonic
dispersion relation leads to competing wave numbers. This phase diagram is
intented to provoke more quantitative theoretical studies, which might
also lead to an understanding of the novel wave phenomena presented here,
namely the oscillating defects and the relaxational phase oscillations.

%\section{Acknowledgment} 
%======================== 
It is a pleasure to thank V.~Frette, H.~W.~M\"uller, R.~Richter, H.~Riecke, 
and M.~I.~Shliomis for helpful discussions. The experiments are
supported by the 'Deutsche Forschungsgemeinschaft' through Re588/10.

\end{document}